\documentclass[conference]{IEEEtran}
\IEEEoverridecommandlockouts
\usepackage{cite}
\usepackage{amsmath,amssymb,amsfonts,dsfont}
\usepackage{algorithmic}
\usepackage{graphicx}
\usepackage{textcomp}
\usepackage{xcolor}
\usepackage{subfigure}

\newcommand{\Fig}[1]{Figure~\ref{fig:#1}}

\newcommand{\Sec}[1]{Sec.~\ref{sec:#1}}

\newcommand{\Eq}[1]{(\ref{eq:#1})}

\begin{document}

\title{
Performance and EMF Exposure Trade-offs\\in Human-centric Cell-free Networks
}

\author{\IEEEauthorblockN{Francesco Malandrino}
\IEEEauthorblockA{CNR-IEIIT \\
Italy}
\and
\IEEEauthorblockN{Emma Chiaramello}
\IEEEauthorblockA{CNR-IEIIT \\
Italy}
\and
\IEEEauthorblockN{Marta Parazzini}
\IEEEauthorblockA{CNR-IEIIT \\
Italy}
\and
\IEEEauthorblockN{Carla Fabiana Chiasserini}
\IEEEauthorblockA{Politecnico di Torino and CNR-IEIIT \\
Italy}
}

\maketitle

\begin{abstract}
In cell-free wireless networks, multiple connectivity options and technologies are available to serve each user. Traditionally, such options are ranked and selected solely based on the network performance they yield; however, additional information such as electromagnetic field (EMF) exposure could be considered. In this work, we explore the trade-offs between network performance and EMF exposure in a typical indoor scenario, finding that it is possible to significantly reduce the latter with a minor impact on the former. We further find that surrogate models represent an efficient and effective tool to model the network behavior.
\end{abstract}

\begin{IEEEkeywords}
cell-free networks; electromagnetic field exposure.
\end{IEEEkeywords}

\section{Introduction and related work\label{sec:intro}}

Next Generation (Next-G) mobile networks will include a vast number of points of access (PoAs) capable of serving their users. PoAs using different technologies, e.g., Wi-Fi access points and cellular base stations, will coexist in the same network, while being controlled in a centralized manner. In this scenario, each user can be served by {\em multiple} PoAs, with different performance and quality-of-service; it follows that the location of a user is not anymore the main factor determining which PoA serves it. Indeed, the very notion of cell as an area covered by a single PoA is fading, leading to what are termed {\em cell-free} mobile networks~\cite{zhang2021beyond,beysens2021blendvlc}.

A major feature of cell-free networks is the great flexibility in deciding how, i.e., through which PoA, each user should be served. Effectively making such decisions requires (i) that network architectures include suitable decision-making entities, and (ii) that those entities have access to the necessary information. Both issues are addressed by the very active research field of virtualized radio access networks (vRAN), where softwareized {\em controllers} running on general-purpose hardware make user and resource management decisions, leveraging status reports and logs coming from different parts of the network infrastructure~\cite{ayala2019vrain,garcia2021ran,MLedge_TMC2021}. Timing is always an important consideration in vRAN scenarios; as an example, the very popular open RAN (O-RAN) paradigm includes three classes of controllers for real-time (latency below 1~ms), near-real-time (around 10~ms), and non-real-time (around 1~s) decisions~\cite{garcia2021ran,taleb2020fully}.

A related issue is {\em which data} to consider when making network management decisions, and the {\em objective} to pursue when making them. The traditional approach~\cite{bonati2021intelligence,ayala2019vrain,MLedge_TMC2021} has long been to maximize a network performance metric, e.g., average throughput or end-to-end service latency, using data on, e.g., interference conditions and user location. However, beyond being fast, Next-G networks are expected to be {\em sustainable} and {\em human-centric}; to achieve these goals, additional information must be collected and accounted for when making network management decisions.

Making any technology -- specifically, mobile networks -- {\em human-centric}  does not only imply accounting for the  needs and preferences of individual users, but also to evaluate the  impact of communication networks on human health. The interactions between electromagnetic fields (EMF) and biological tissues are indeed related to a wide range of biological effects, depending both on the frequency content and EMF amplitude. Concerning radiofrequency EMF (RF-EMF), the only scientifically recognized and proven effect of exposure is heating of biological tissues. The International Commission on Non-Ionizing Radiation Protection (ICNIRP) guidelines have defined the limits of exposure to RF-EMF as thresholds below which RF-EMF exposure is safe according to scientific knowledge~\cite{international2020guidelines}. %
Nonetheless, the development of new technologies requires an evaluation of the levels of exposure, as one fundamental step for proper health risk assessment process.  The health risk management aspects, as well as the current public concern on possible long-term health effects caused by exposure to new RF-EMF emitting sources~\cite{chiaraviglio2021cellular}, even if at levels below guidelines, make the EMF exposure evaluation playing an important role in new generation technologies.

In spite of the vast research efforts towards Next-G networks~\cite{martin2021kpi,malandrino2019reducing}, as well as  on EMF exposure assessment on human health 
\cite{chiaramello2019stochastic}, research about EMF-aware network {\em management} is surprisingly scarce. Indeed, most existing works focus on the network {\em planning} stage, e.g., placing the PoAs in such a way that EMF exposure targets (or limits) are not exceeded~\cite{chiaraviglio2018planning,chiaraviglio2021cellular}. These approaches, however, fail to exploit the capability of modern virtualized, software-defined, cell-free networks.

In this work, and unlike prior art, we account for an EMF exposure  at network {\em management} time, and envision making decisions to constantly adapt the network configuration to the external conditions and user demand, thereby ensuring adequate performance while keeping EMF exposure low. Our work is thus a contribution towards the higher-level goal of making the network as a whole more {\em human-centric}, hence sustainable.

To this end, a first, concrete goal is to characterize the  {\em trade-offs} existing between performance and EMF exposure levels, hence, to obtain a preliminary understanding on whether it is feasible to significantly reduce the latter without jeopardizing the former. Network performance is assessed through simulation, considering a typical office layout, mmWave technology, and the experiment-based model presented in~\cite{ju2021millimeter}.
As for the EMF exposure assessment, we apply a stochastic approach, based on the use of surrogate models accounting for the uncertainty coming from users' positions. Thanks to their relative simplicity, surrogate models based on polynomial basis~\cite{chiaramello2019stochastic} can be a building block of {\em lightweight} solution strategies, appropriate for decisions to make under tight time constraints and over hardware with limited capabilities.

The remainder of this paper is organized as follows. We begin, in \Sec{scenario}, by presenting the reference scenario we consider for this work. We then detail the methodology we use in \Sec{methodology}, before presenting our numerical results in \Sec{results}. We discuss the main take-away messages in \Sec{discussion}, along with open directions for future research. Finally, we conclude the paper in \Sec{conclusion}.

\section{Reference scenario\label{sec:scenario}}

\begin{figure}
\centering
\includegraphics[width=.8\columnwidth]{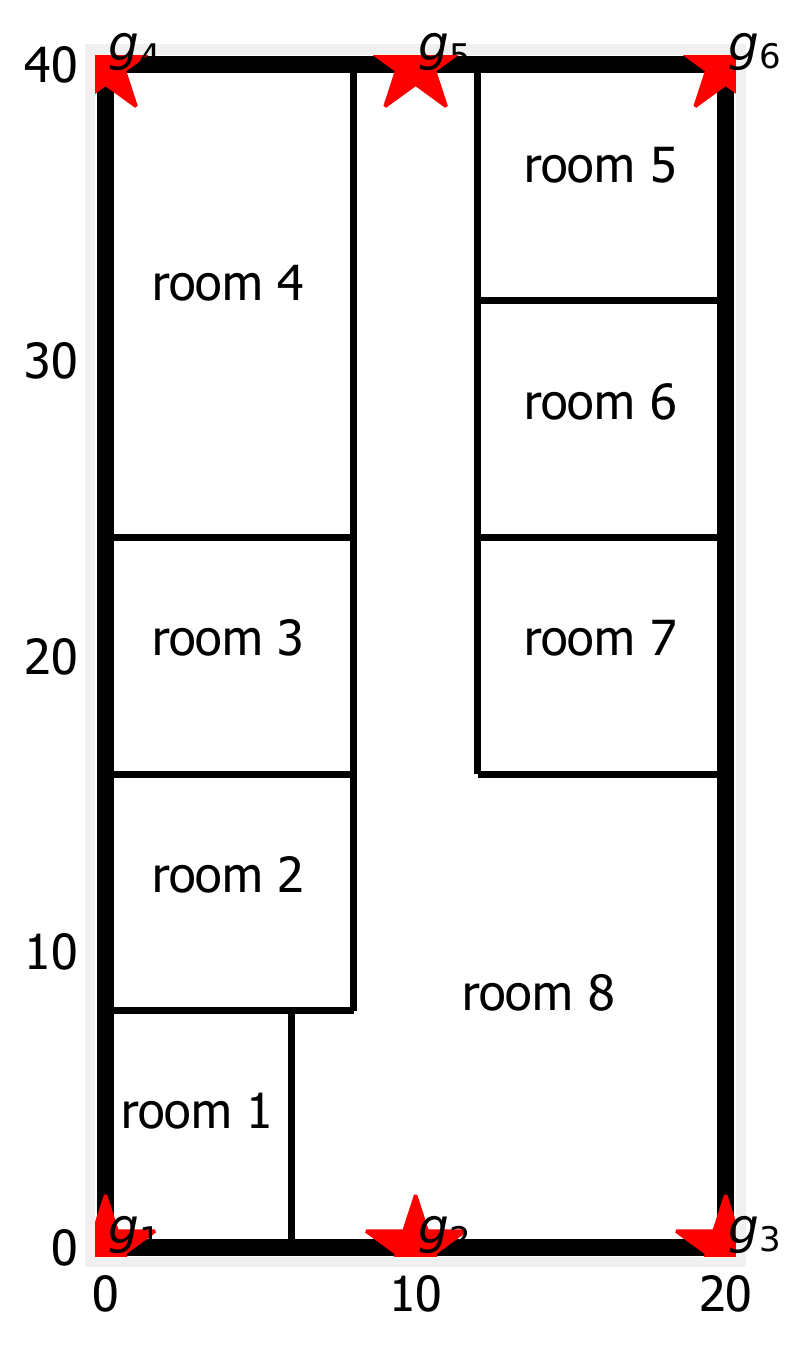}
\caption{
Our reference topology, with red stars representing gNBs and the number of each room indicated therein. Coordinates are in meters.
\label{fig:office}
} %
\end{figure}

We focus on a typical indoor scenario, specifically, the office layout depicted in \Fig{office}, similar to the one used in many works on indoors EMF exposure~\cite{tognola2021use}.

The office contains a total of six PoAs, namely, mmWave gNBs, called~$g_1\dots g_6$, and represented by red stars in \Fig{office}. As in~\cite{ju2021millimeter}, gNBs operate at a frequency of 28~GHz, and their transmission power is set to 23.9~dBm (245~mW). gNBs emit {\em beams} whose width is~$15^{\circ}$, and the path loss incurred is given by~\cite{ju2021millimeter}:
\begin{equation}
\label{eq:rappaport}
\text{PL}_{\text{dB}}(f,d)=20\log_{10}\left(\frac{4\pi d_0 f}{c}\right)+10n\log_{10}\left(\frac{d}{d_0}\right).
\end{equation} 
In \Eq{rappaport}, $d$~is the distance between  transmitter and  receiver, $d_0$=1~m is the reference distance, $f$=28~GHz is the frequency, and $c$~is the speed of light. $n$ is the path-loss exponent, which also accounts for whether there are obstacles, e.g., walls, between transmitter and receiver (if there are no obstacles, we are in  line-of-sight (LoS) conditions). In our case, following the experiments reported in~\cite{ju2021millimeter}, we set~$n=1.7$ for LoS conditions, and~$n=4.6$ for non-LoS conditions.

A total of ten users are randomly placed across the topology. In pure cell-free fashion, any gNB could serve any of the rooms; in the following, we will refer to a gNB-to-room assignment as a {\em strategy}. We will compare a total of~32 such strategies over 1,000~different user placements, and study the average {\em and distribution} of the network performance (expressed as the achievable data rate) and
EMF exposure (expressed as power density, consistently with the reference levels provided by the ICNIRP guidelines~\cite{international2020guidelines}). 
Importantly, network performance is evaluated at the users' locations, while EMF exposure is evaluated across the whole topology.

\section{Methodology\label{sec:methodology}}

\begin{figure*}
\centering
\includegraphics[width=\textwidth]{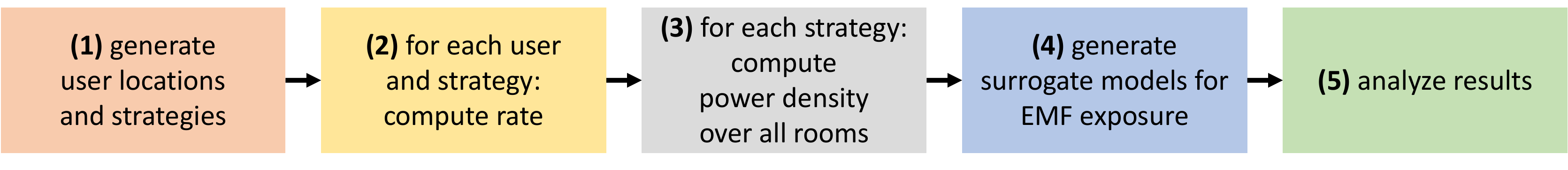}
\caption{
The main steps of our methodology.
\label{fig:flow}
} %
\end{figure*}

Our methodology includes five main steps, as summarized in \Fig{flow}.

The first step, {\bf (1)} in \Fig{flow}, is to generate user positions and network management strategies. Concerning the former, we generate a total of 1,000 {\em scenarios}, all using the office layout in \Fig{office} and each including 10 users. Each room contains a user, except rooms~4 and~8 which contain two each; in each scenario, users are randomly placed within their rooms. The distribution of the coordinates describing the positions of the users in each room is assumed to be uniform, and scenarios were obtained through the Latin hypercube sampling strategy~\cite{blatman2007quasi}.

\begin{figure}
\centering
\includegraphics[width=\columnwidth]{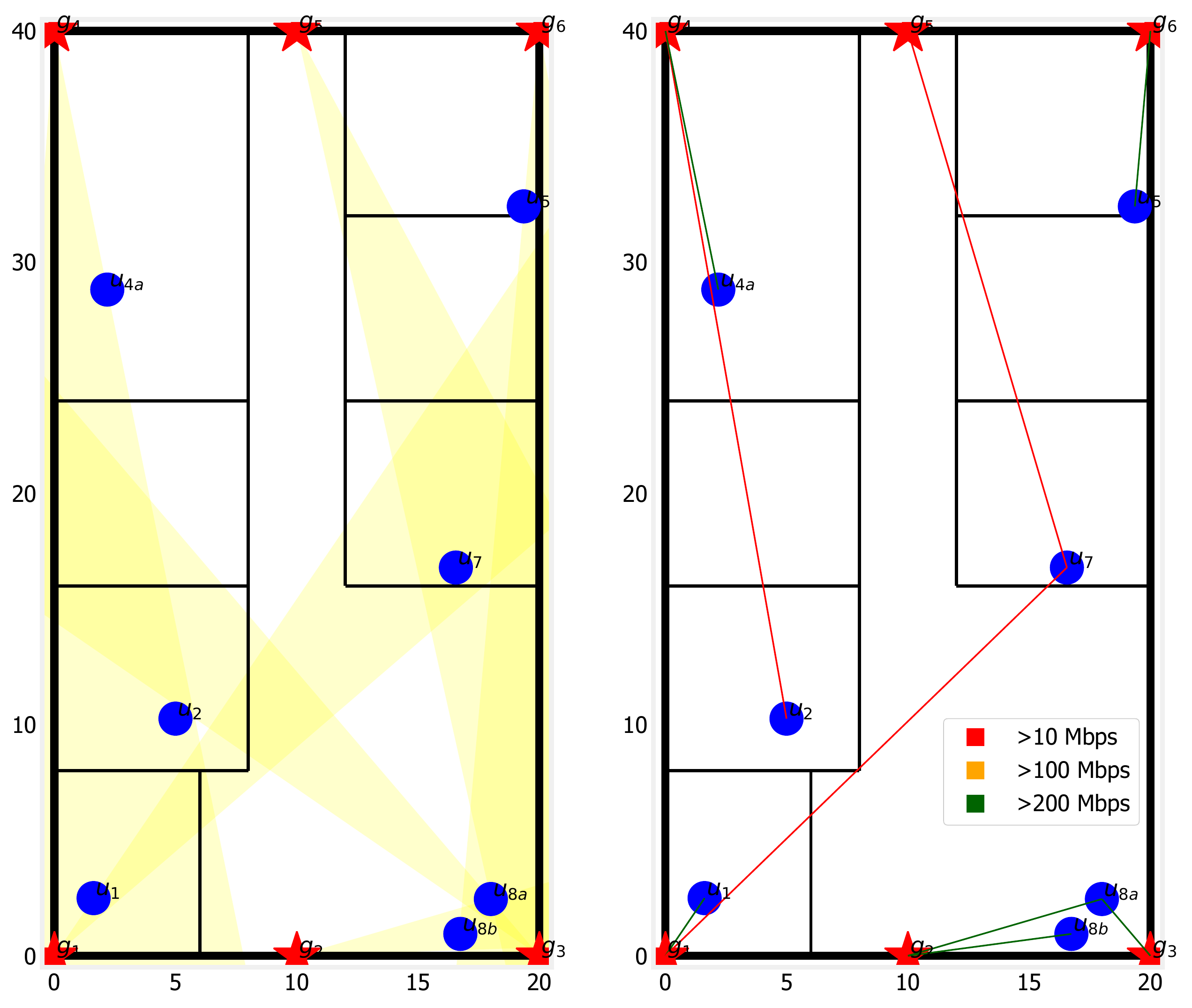}
\caption{
An example network configuration: beam directions (left) and resulting data rate (right).
\label{fig:example-configuration}
} %
\end{figure}

As for network management strategies, we assume that each gNB is pointed towards a room, and identify a {\em strategy} as a gNB-to-room association. As an example,  \Fig{example-configuration} depicts a strategy where gNBs $g_1$--$g_6$ are pointed, respectively, towards rooms $1$, $8$, $8$, $4$, $8$, and $5$. Notice that where a gNB -- more precisely, its beam -- is {\em pointed} does not limit the users it can serve: in the example in \Fig{example-configuration}, gNB $g_1$ points towards room~$1$, but it also serves room~$7$. Similarly, nothing prevents multiple gNBs from pointing their beams towards the same room.

In step {\bf (2)}, we leverage the model and results from~\cite{ju2021millimeter} to assess the network performance and behavior. Specifically, we apply \Eq{rappaport} to determine the signal-to-noise-and-interference-ratio (SINR) experienced by each user, and then computing the achievable rate via Shannon's formula $\text{rate}=B\log_2(1+\text{SINR})$,
where $B=500$~MHz is the available bandwidth. In the example of \Fig{example-configuration}, it is possible to notice how longer distances and/or obstacles (hence, non-LoS conditions) between the gNB and the user result in lower rates.

In step {\bf (3)}, we move to EMF exposure. Unlike data rate, we are interested in the exposure over the whole topology, as there may be other people in it (e.g., visitors) which are not users of the network. Therefore, we superimpose to the office layout a regular {\em grid} made of $100\times 200$ points, spaced $20$~cm apart from each other. For each point of the grid, we use \Eq{rappaport} to compute the total received power (from all gNBs). Then, we normalize such power to the actual area of the receiver to obtain the power density. The $95^{\text{th}}$ percentile ($S_{95}$) and the mean value ($S_{\text{mean}}$) of the power density values obtained in each scenario and for each network management strategies were considered as EMF exposure metrics. 

In step {\bf (4)}, we build {\em surrogate models} for EMF exposure, based on Low Rank Tensor Approximations (LRA)~\cite{konakli2016polynomial}. In general, EMF exposure could be expressed as $Y = M(X)$ were $Y$~represents the exposure metrics (in this study, $S_{95}$ and $S_{\text{mean}}$) and $X$~represents the variables known to influence the exposure scenarios (i.e., the positions of the users in our case). Such quantities are linked by an unknown, and potentially very complex, model~$M$, i.e.,
\begin{equation}
\nonumber
Y=M(X)\,.
\end{equation}
Then, a surrogate model~$\tilde{M}$ is an approximation of the original model, showing similar statistical properties but significantly lower mathematical complexity, allowing us to write:
\begin{equation}
\nonumber
Y\approx\tilde{M}(X)\,.
\end{equation}

Building a surrogate model requires a finite set of observations of the original model, called {\em experimental design}; in this study, the experimental design is the set of 1,000 scenarios with different users locations. Different methods can be used to build surrogate models; LRA belongs to the class of {\em non-intrusive} methods, as it does not require any additional information on the phenomenon being studied, which is treated as a ``black box''. It aims at developing surrogate models containing polynomial functions in high dimensional spaces based on canonical decomposition. A representation of the surrogate model as a finite sum of rank-one functions reads as:
\begin{equation}
\label{eq:LRA1}
Y_{\text{LRA}} = \sum_{l=1}^{R}{b_lw_l} = \sum_{l=1}^{R}{b_l \prod_{i=1}^{M}{v_l^{(i)}X_i}}\,,         
\end{equation}
where $w_l$ is the {\em l}-th rank-one function obtained as product of univariate functions of the components of $X_i$, $v_l^{(i)}$ denotes a univariate function of the components of $X_i$ in the {\em l}-th rank-one component, {\em M} is the number of input variables, $b_l, (l = 1,\dots, R)$ are scalars that can be viewed as normalizing constants, and $r$ is the rank of the decomposition.

By exploiting the tensor-product structure of the multivariate polynomial basis, as suggested by~\cite{konakli2016polynomial}, and expanding $v_l^{(i)}$ into a polynomial basis that is orthonormal with respect to the marginal distribution of the input parameters $X_i$, \Eq{LRA1} can be transformed into:
\begin{equation}
\label{eq:LRA2}
Y_{\text{LRA}} = \sum_{l=1}^{R}{b_l \prod_{i=1}^{M}{\sum_{k=0}^{p_i}{z_{k,l}^{(i)}P_{k}^{(i)}(X_i)}}}\,,           
\end{equation}
where $P_k^{(i)}$ denotes the $k$-th degree univariate polynomial in the $i$-th input variable, $p_i$ is the maximum degree of $P_k^{(i)}$ and $z_{k,l}^{(i)}$ is the coefficient of $P_k^{(i)}$ in the $l$-th rank-one component. 

The choice of the proper polynomial basis $P_k^{(i)}$ that would be used to build up the LRA model is based on the criteria of orthonormality to the marginal distributions of the input parameters $X_i$: as the input parameters $X_i$ were supposed to be uniformly distributed, in this study the Legendre polynomials have been used.
In order to estimate the unknown parameters, i.e., the polynomial coefficients $z_{k,l}^{(i)}$ and the normalizing coefficients $b_l, (l = 1,\dots, R)$ of the surrogate model, we have adopted a greedy algorithm \cite{konakli2016polynomial,chevreuil2015least}, based on Alternated Least-Squares (ALS) minimization. ALS is predicated on sequentially updating the coefficients along separate dimensions, and progressively increasing their rank by successively adding rank-one components. The employed algorithm involves alternating {\em correction} steps and {\em updating} steps. In the $r$-th correction step, the rank-one tensor $w_r$ is built, while in the $r$-th updating step the set of normalizing coefficients $b_1,\dots, b_r$ is determined. In order to obtain sparse low rank approximations, the approach described by~\cite{konakli2016polynomial} has been integrated by solving all the minimization problems using the hybrid least angle regression method~\cite{blatman2007quasi,efron2004least}. This proposed approach  permits to obtain sparsity both for each rank-one tensor $w_r$, discarding non-significant polynomials, and for the complete $Y_{\text{LRA}}$ model, discarding non-significant rank-one tensors (for more details, the interested reader is referred to~\cite{chiaramello2019children}).

To select the best rank $r$ for the LRA model, we have applied the method proposed in~\cite{chevreuil2015least}, based on a 3-fold cross validation. The experimental design was divided into three subsets and, iteratively, three LRA models were built considering two among the three subsets as training set. For each model, the root mean square error between the values estimated with the LRA model and those of the respective testing set was estimated. The rank $r$ yielding the smallest average root mean square error over the three LRA models was identified as optimal. Then, a new LRA model of rank $r$ was built using the full experimental design. The 3-fold error, obtained averaging the mean square errors calculated on each training set and normalized on the empirical variance of the set provided a fair approximation of the generalized error of the surrogate models.
Once obtained reliable surrogate models of the exposure, we used them to estimate the probability density functions of $S_{95}$ and $S_{\text{mean}}$ for each considered strategy. 

Finally, step {\bf (5)} of our methodology consists of analyzing the resulting data, presented in \Sec{results} next.

\section{Numerical results\label{sec:results}}

\begin{figure}
\centering
\includegraphics[width=\columnwidth]{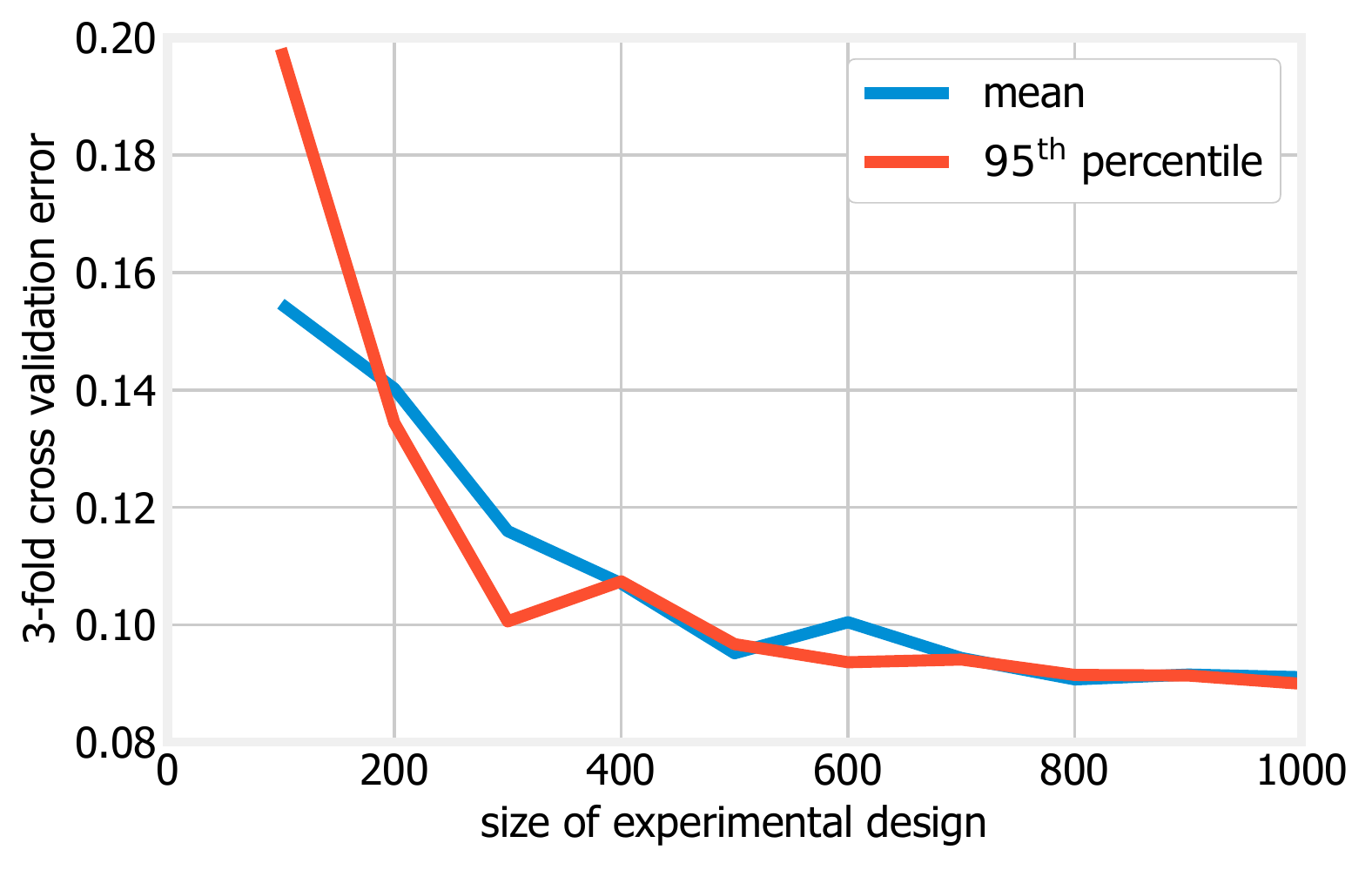}
\caption{
3-fold cross validation error obtained in building LRA based surrogate models for $S_{95}$ and $S_{\text{mean}}$ for increasing sizes of experimental design (network management strategy no.\,9).
\label{fig:3fold}
} %
\end{figure}

A first issue we seek to investigate concerns how effective the surrogate models are, i.e., how well they can represent the real behavior of the network. To this end,
\Fig{3fold} shows the 3-fold cross validation error describing the accuracy of LRA surrogate models in estimating $S_{95}$ and $S_{\text{mean}}$ for a specific network management strategy (namely, strategy no.\,9), as size of the experimental design (i.e., the number of considered exposure scenarios) grows. As expected, when increasing the size of the experimental design, the 3-fold cross validation error decreases. For this particular strategy, an experimental design larger than 500 was enough to obtain errors lower than 0.1 for LRA estimation of both $S_{95}$ and $S_{\text{mean}}$. Across the different strategies, we chose sizes of the experimental design large enough to built surrogate models with 3-fold cross validation error lower than 0.1, for both $S_{95}$ and $S_{\text{mean}}$. 

\begin{figure}
\centering
\includegraphics[width=\columnwidth]{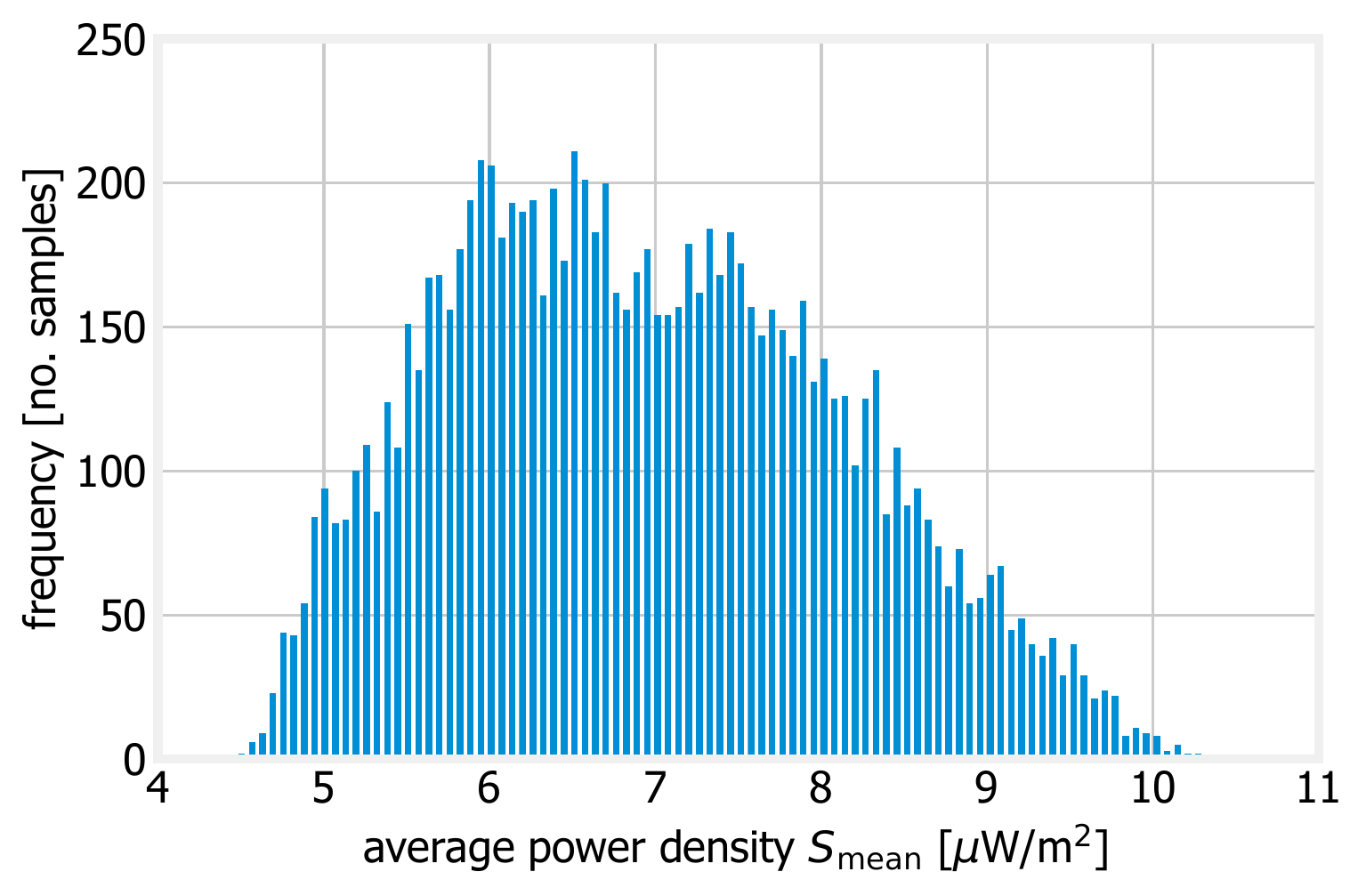}
\caption{
Histograms of $S_{\text{mean}}$ obtained for 10,000 exposure scenarios when considering network management strategy no.\,9.
\label{fig:HIST}
} %
\end{figure}

Once we have the LRA surrogate models, we generate a set of 10,000 exposure scenarios, and compute $S_{95}$ and $S_{\text{mean}}$ through the surrogate models. As an example, \Fig{HIST} shows the histogram of the $S_{\text{mean}}$ values obtained for one network management strategy (namely, no.\,9). The data follows a log-normal distribution, with mean value equal to 6.97 $\mu$W/m$^2$ and slightly positive skewness, equal to 0.29. This highlights that, for most scenarios, $S_{\text{mean}}$ values are lower than the mean value of the distribution.

\begin{figure}
\centering
\includegraphics[width=\columnwidth]{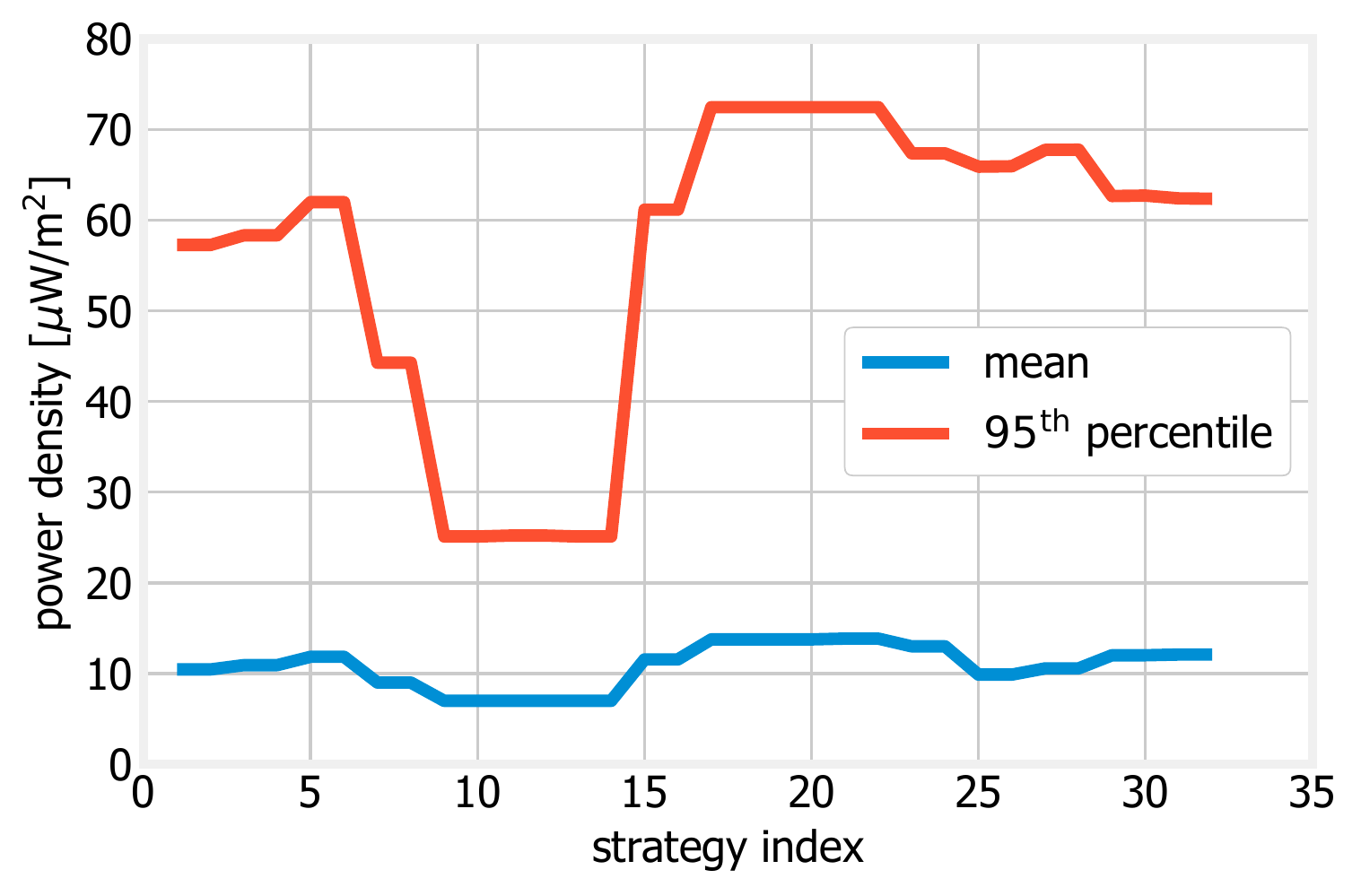}
\caption{
Mean values of $S_{95}$ and $S_{\text{mean}}$ across network management strategies.
\label{fig:EMFexp}
} %
\end{figure}

Next, we seek to assess how much EMF exposure varies across strategies; intuitively, this is linked to whether or not it makes sense to even {\em seek} for low-exposure strategies. To this end, \Fig{EMFexp} shows the mean values of $S_{95}$ and $S_{\text{mean}}$ estimated by LRA surrogate models for each considered strategy. It is possible to observe a very significant variability in terms of $S_{95}$: strategies from no.\,1 to 6 show $S_{95}$ equal to about 60\,$\mu$W/m$^2$, strategies no.\,7 and 8 show $S_{95}$ equal to about  40\,$\mu$W/m$^2$, while strategies from no.\,9 to 14 show the lowest $S_{95}$ values, equal to about 25\,$\mu$W/m$^2$. All the remaining strategies show $S_{95}$ values higher than 60\,$\mu$W/m$^2$, with the highest values observed for strategies from no.\,17 to 22. As for the $S_{\text{mean}}$ values, even if a dependence from the network management strategies can be observed, the obtained values are all in the range 7-14\,$\mu$W/m$^2$. Importantly, all such values are well below the reference level
indicated by the ICNIRP guidelines \cite{international2020guidelines}.   
\begin{figure}
\centering
\includegraphics[width=\columnwidth]{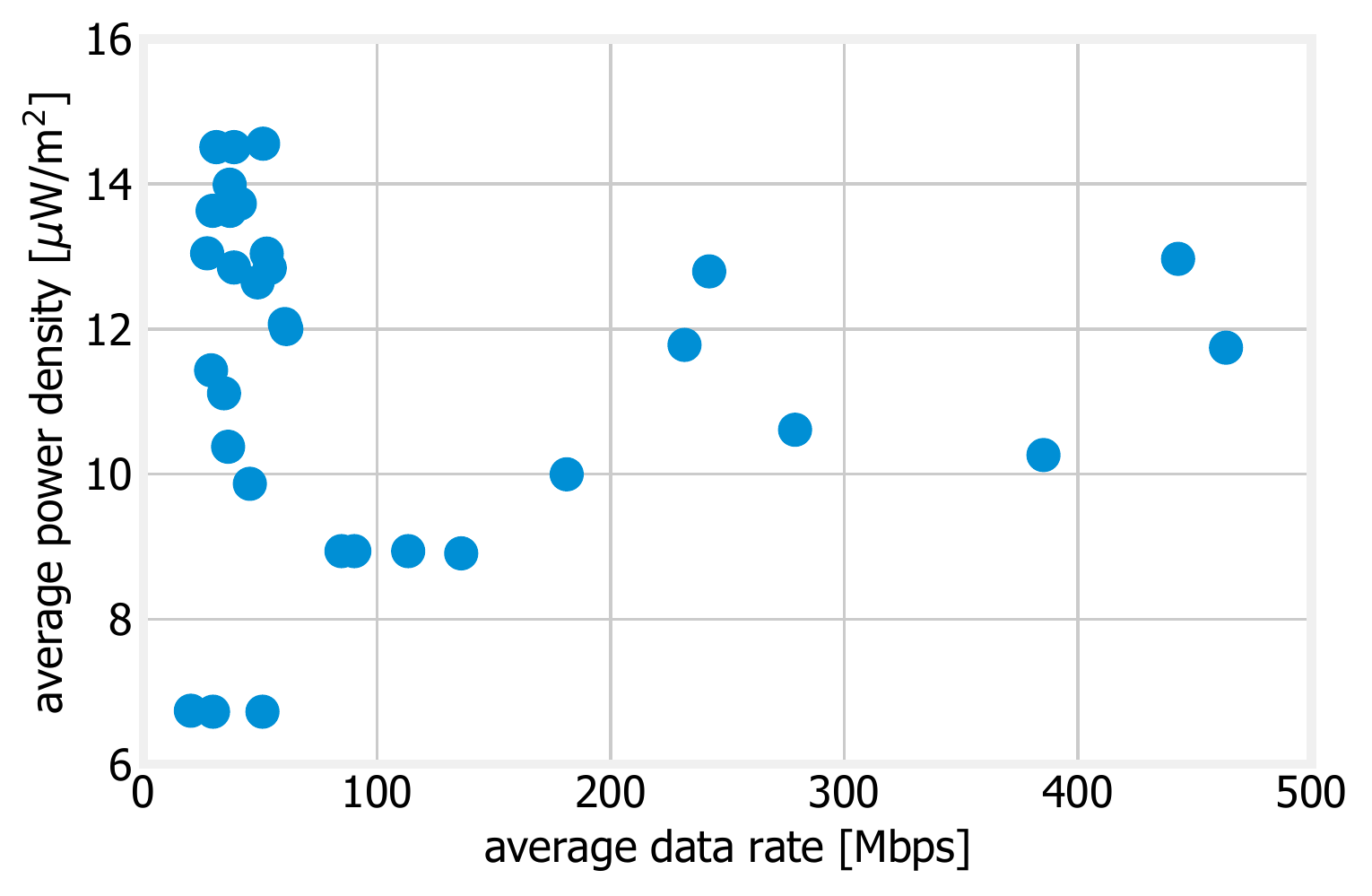}
\caption{
Network performance and EMF exposure under different strategies. Each marker corresponds to one strategy, and its positions over the $x$- and $y$-axes correspond, respectively, to the average data rate and the power density.
\label{fig:scatter}
} %
\end{figure}

Last, we characterize the possible trade-offs between EMF exposure and network performance, summarized in \Fig{scatter}. In the plot, each marker corresponds to a different strategy; the location of the marker along the $x$- and $y$-axes corresponds to, respectively,  the network performance (expressed as the average data rate) and EMF exposure (expressed through the power density, computed in step~3 as per \Sec{methodology}).

Focusing on the rightmost part of the plot, we can observe that there are multiple strategies resulting in very good network performance, very likely exceeding the requirements of the users. Even more importantly, such strategies result in very different levels of EMF exposure, consistently with \Fig{EMFexp}. In other words, it is possible to substantially reduce EMF exposure -- by over~30\% -- without any noticeable change to the network performance. Whether or not this is necessary or advisable depends upon the concrete scenario at hand; however, the results in \Fig{scatter} confirm our intuition that non-trivial trade-offs between EMF exposure and network performance do exist and are worth exploring and exploiting.

\section{Discussion and open issues\label{sec:discussion}}

The numerical results  presented in \Sec{results} show that interesting, high-quality trade-offs between network performance and EMF exposure do exist and are worth exploring. Furthermore, LRA based surrogate models are a viable tool to model and predict the behavior of cell-free networks with a low complexity. Building upon such take-away messages, it is possible to identify several promising research directions to explore.

A first one concerns {\em energy efficiency}, which is tightly related to the issue of sustainability. As \Fig{scatter} suggests, it is possible to obtain network configurations that have a high performance {\em and} a low EMF exposure. Interestingly, such solutions tend to {\em also} be more energy-efficient: indeed, interference among gNBs is the main factor degrading network performance, and gNBs needlessly emitting power also increase EMF exposure. It follows that including gNB transmission power among the decisions to make, and energy efficiency among the metrics to account for, can yield further performance gains for the network, whilst also furthering the objectives of sustainable, human-centric networking.

A second major research avenue concerns {\em how} to make network management decisions. In our simulations  we compared a set of pre-existing possible strategies; however, in real-world conditions strategies must be built on the fly, and updated according to quickly-changing external conditions. This requires decision-making approaches that are both effective (i.e., they yield good-quality decisions) and efficient (i.e., they reach such decisions swiftly). These two requirements tend to contradict one another, hence, a trade-off between the two must be sought. In this context, accounting for additional aspects, such as EMF exposure and energy efficiency, increases the quality of decisions, at the cost of -- potentially -- rendering them complex to make.

Regardless of {\em how} decisions are made, we will always need a way to estimate the effect of the network configuration on the performance and behavior of the network. To this end, the surrogate model approach we have demonstrated represents a viable alternative to more popular machine learning approaches like deep neural networks (DNNs)~\cite{bonati2021intelligence,ayala2019vrain,MLedge_TMC2021}. Compared to DNNs, surrogate models have two main advantages, namely:
\begin{itemize}
    \item determining the model coefficients (i.e., training the machine learning models) requires less data (as per \Fig{3fold}) and is less computationally intensive;
    \item using the model (i.e., the inference phase) simply requires computing a polynomial \Eq{LRA2}, which is more resource-efficient than running through a DNN.
\end{itemize}
This is especially important in virtualized scenarios~\cite{bonati2021intelligence,ayala2019vrain}, where learning has to take place within the network nodes itself, hence, with limited resources.

Finally, the sustainability of machine learning itself is a hotly debated issue~\cite{schwartz2020green}. Techniques allowing to complement -- or, in some cases, even dispense with -- DNNs can therefore have a significant impact, even beyond applications to networking.

\section{Conclusion\label{sec:conclusion}}

We have considered the scenario of cell-free networks, where multiple possible network management strategies exist, leveraging different PoAs to serve the users. As a part of the overarching general push towards {\em human-centric} networking, we have envisioned including EMF exposure in the network management process, seeking for management strategies combining high performance and low EMF exposure levels.

To validate our intuition, we have considered a typical office layout with six gNBs operating at 28~GHz, and used surrogate models based on LRA to model EMF exposure through the mean and $95^{\text{th}}$~percentile of the power density. We have found that multiple, high-quality trade-offs between network performance and EMF exposure can be explored, hence, it is worth including exposure considerations in the network management process. Furthermore, LRA surrogate models were remarkably effective in estimating EMF exposure, which makes them a promising, {\em lightweight} tool to leverage for decision making in scenarios with tight deadlines and/or limited hardware capabilities.

Future research directions include defining decision-making algorithms, combining EMF exposure with energy efficiency, and exploring the extent to which surrogate models can complement DNNs.

\section*{Acknowledgement}

This work was supported by the NPRP-S 13th Cycle Grant No.\,NPRP13S-0205-200265 from the Qatar National Research Fund (a member of Qatar Foundation).  The views expressed are those of the authors and do not necessarily represent the project.

\bibliographystyle{IEEEtran}
\bibliography{refs}

\end{document}